\documentclass[prx,twocolumn,floatfix,superscriptaddress,amsmath,aps,longbibliography]{revtex4-1}
\pdfoutput=1
\usepackage{dcolumn,graphicx,color,booktabs,microtype,afterpage,bm}
\makeatletter\renewcommand{\fnum@figure}[1]{\figurename~\thefigure.}\makeatother
\makeatletter\renewcommand{\fnum@table}[1]{\tablename~\thetable.}\makeatother
\usepackage[colorlinks,plainpages=false,linkcolor=blue,urlcolor=blue,citecolor=blue,pdfpagemode=UseNone,pdfstartview=FitBH]{hyperref}
\newcommand{\ste}{NaMgFe(C$_2$O$_4$)$_3\cdot$9H$_2$O}

\begin{document}

\title{Direct determination of the zero-field splitting for Fe$^{3+}$ ion in a synthetic polymorph of the oxalate mineral stepanovite NaMgFe(C$_2$O$_4$)$_3\cdot$9H$_2$O: a natural MOF}

\author{Tao~Xie}
\affiliation{Neutron Scattering Division, Oak Ridge National Laboratory, Oak Ridge, Tennessee 37831, USA}
\author{S.~E.~Nikitin}
\thanks{Present address: Paul Scherrer Institute, Villigen PSI CH-5232, Switzerland}
\affiliation{Max Planck Institute for Chemical Physics of Solids, D-01187 Dresden, Germany}
\affiliation{Institut f{\"u}r Festk{\"o}rper- und Materialphysik, Technische Universit{\"a}t Dresden, D-01069 Dresden, Germany}
\author{A.~I.~Kolesnikov}
\affiliation{Neutron Scattering Division, Oak Ridge National Laboratory, Oak Ridge, Tennessee 37831, USA}
\author{E.~Mamontov}
\affiliation{Neutron Scattering Division, Oak Ridge National Laboratory, Oak Ridge, Tennessee 37831, USA}
\author{L.~M.~Anovitz}
\affiliation{Chemical Sciences Division, Oak Ridge National Laboratory, Oak Ridge, Tennessee 37831, USA}
\author{G.~Ehlers}
\affiliation{Neutron Technologies Division, Oak Ridge National Laboratory, Oak Ridge, TN 37831, USA}
\author{I. Huski\'{c}}
\affiliation{Department of Chemistry, McGill University, 801 Sherbrooke Street West, Montreal, H3A 0B8 Quebec, Canada.}
\author{T. Fri\v{s}\v{c}i\'{c}}
\affiliation{Department of Chemistry, McGill University, 801 Sherbrooke Street West, Montreal, H3A 0B8 Quebec, Canada.}
\author{A.~Podlesnyak}
\thanks{Corresponding author: podlesnyakaa@ornl.gov}
\affiliation{Neutron Scattering Division, Oak Ridge National Laboratory, Oak Ridge, Tennessee 37831, USA}

\begin{abstract}
We employed inelastic neutron scattering (INS), specific heat, and magnetization analysis to study the magnetism in a synthetic polymorph of the quasi-two-dimensional natural metal-organic framework material, stepanovite \ste.  No long-range magnetic order can be observed down to 0.5 K. The INS spectra show two dispersionless excitations at energy transfer 0.028(1) and 0.050(1)~meV at base temperature, which are derived from the magnetic transitions between zero-field splitting (ZFS) of $S$~$=5/2$ ground state multiplets of Fe$^{3+}$ ion. Further analysis of the INS results shows that the Fe$^{3+}$ ion has an easy-axis anisotropy with  axial ZFS parameter $D$~=~$-0.0128(5)$~meV and rhombic parameter $E$~=~0.0014(5)~meV. The upward behavior at zero field and Schottky-like peak under magnetic field of the low-temperature magnetic specific heat further support the INS results. Our results clearly reveal the magnetic ground and excited state of this stepanovite polymorph.
\end{abstract}

\maketitle{}

\section{Introduction}
\label{Intro}

Metal-organic frameworks (MOFs) are a recently-emerged  family of porous crystalline materials, with abundant topological structures, that have broad prospects for applications in gas storage and separation, carbon sequestration, catalysis, photovoltaics, biomedicine delivery, and chemical sensing, etc~\cite{Huskic2016,Kirchon}. The oxalate-based organic minerals zhemchuzhnikovite and stepanovite, \ste, have an open structure that is analogous to MOF materials that have been developed by several groups, making them so far rare examples of naturally-formed MOF architectures~\cite{Huskic2016}.
Although metal oxalates are the largest family of organic minerals \cite{Echigo}, natural zhemchuzhnikovite and stepanovite crystals are exceptionally rare, found only in exotic geochemical environments.
Recently, however, different research groups, including ours, have shown that synthetic stepanovite crystals (structurally and morphologically identical to the natural ones) can be readily obtained in the form of millimeter-sized, light green triangular crystals~\cite{Huskic2016,Piro2016}.
Stepanovite crystallizes in a trigonal structure (space group $R3c$) with
lattice parameters $a={9.887}(13)$~{\AA} and $c={37.03}(5)$~\AA ~\cite{Knipovich,Huskic2016,Piro2016}.

\begin{figure}[tb]
\center{\includegraphics[width=1\linewidth]{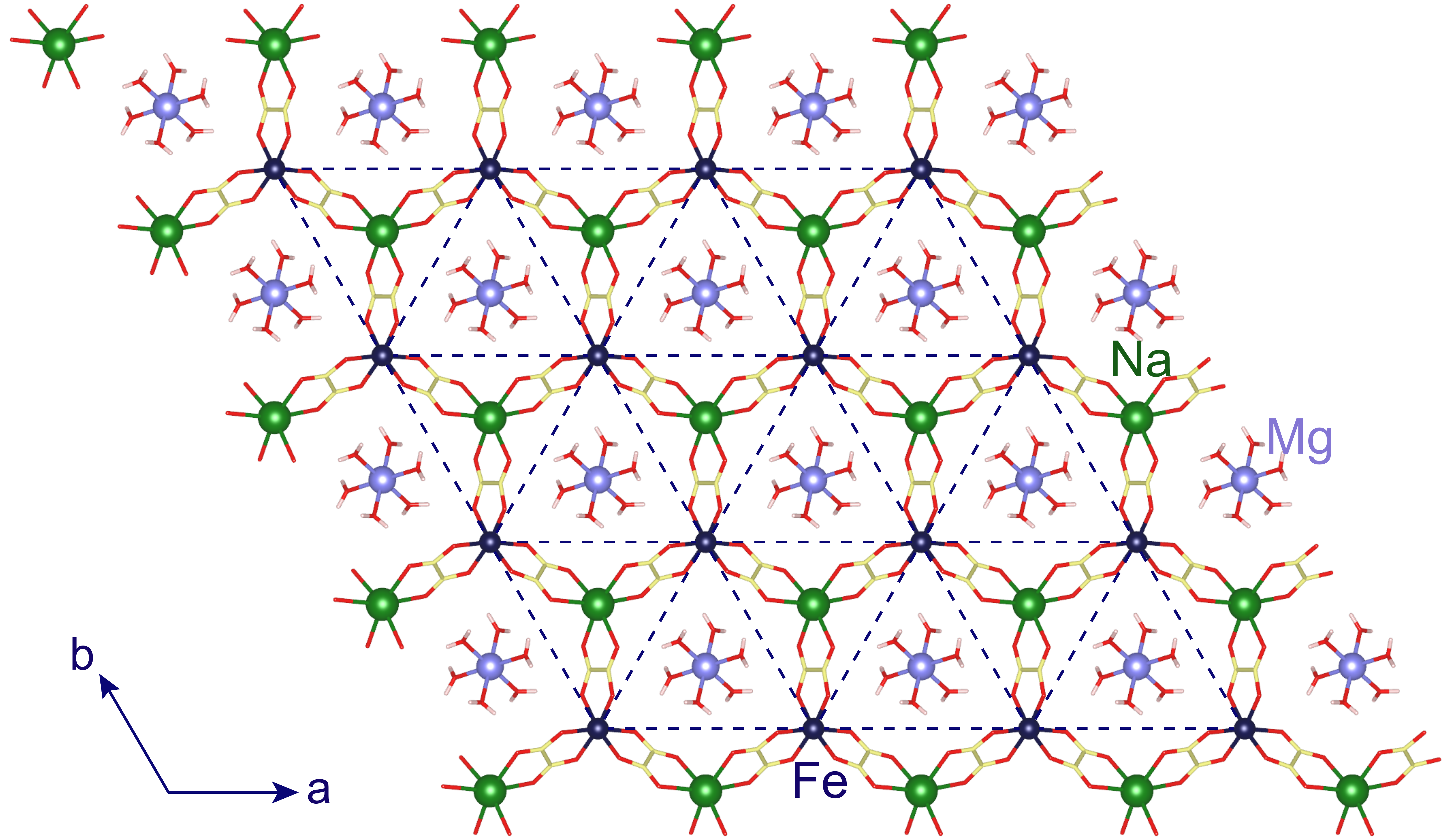}}
  \caption{~Schematic view of a single NaMgFe(C$_2$O$_4$)$_3$ layer, viewed along the crystallographic $c$-axis. The triangular Fe$^{3+}$ sublattice is highlighted as blue broken lines.
  }
  \label{structure}
\end{figure}

It was recently shown that the laboratory synthesis of stepanovite, depending on the synthetic conditions, can also lead to the formation of stepanovite II, a polymorph isostructural to the mineral zhemchuzhnkovite, which crystallizes in the trigonal space group $P3c1$ with
lattice parameters $a={17.033}(11)$~{\AA} and $c={12.4160}(8)$~\AA~\cite{Huskic2019}.
Structures of both stepanovite forms are dominated by honeycomb NaMgFe(C$_2$O$_4$)$_3$ layers (see Fig.~\ref{structure}), which stack in the ABCABC fashion in stepanovite, and in the stepanovite II adopt the ABAB arrangement.
In each layer, the Fe$^{3+}$ ions are in an octahedral coordination, with Fe-O bonds about 2.03~\AA\ long.
The Fe$^{3+}$ ions can further form a triangular sublattice in the $ab$ plane (see Fig.~\ref{structure}). The Fe-Fe distance in $ab$ plane of stepanovite is equal to the lattice parameter $a={9.887}(13)$~\AA, and in stepanovite II it is 10.028~\AA~\cite{Huskic2019}.
In both structures, adjacent layers are primarily held together through hydrogen bonding to water molecules, and are $\sim$~6.21~\AA\ apart from each other along the $c$-axis, giving rise to a quasi-two-dimensional (2D) structure.

The quasi-2D nature of its layered honeycomb crystal structure and the triangular Fe$^{3+}$ sublattice make \ste\ a favorable material for the display of novel magnetic properties, which led us to speculate whether the Fe$^{3+}$ sublattice might have an ordered/frustrated magnetic ground state or exhibit single-ion properties.
If the spin interactions between Fe$^{3+}$ ions restricted by the low-dimensional structure are strong, one could expect enhanced spin fluctuations to occur~\cite{Moessner}.
Moreover, a geometrically frustrated system, in which the geometry of the lattice prohibits the simultaneous minimization of all interactions can demonstrate exotic phenomena such as spin ice or spin liquid phases.
Recent studies of a growing number of newly synthesized 2D frustrated magnetic materials open a world of fascinating magnetic phenomena and quantum phase transitions for analysis and application~\cite{Schmidt2017}.

On the other hand, if the Fe-Fe magnetic interactions are negligible, single-ion anisotropy will become the crucial property.
Extensive studies of mononuclear, single-molecular magnets during the last two decades have established that the electronic spin and the crystal field (CEF) splitting of individual metal ions govern the magnetic behavior of such systems~\cite{Woodruff,Meihaus,Bar}.

Given these two options, a natural question is what kind of properties and phenomena exist in stepanovite-type structures?
In that context, as synthetic stepanovite and stepanovite II have been shown to exhibit considerable proton conduction properties~\cite{Sadakiyo}, it is clear that natural MOF materials might exhibit both the structures and properties of advanced laboratory designed materials.
We, therefore, performed inelastic neutron scattering (INS), heat capacity, and  magnetization studies on stepanovite~II in order to determine the magnetic properties of such a quasi-2D MOF material.

\begin{figure}[t!]
\center{\includegraphics[width=1.0\linewidth]{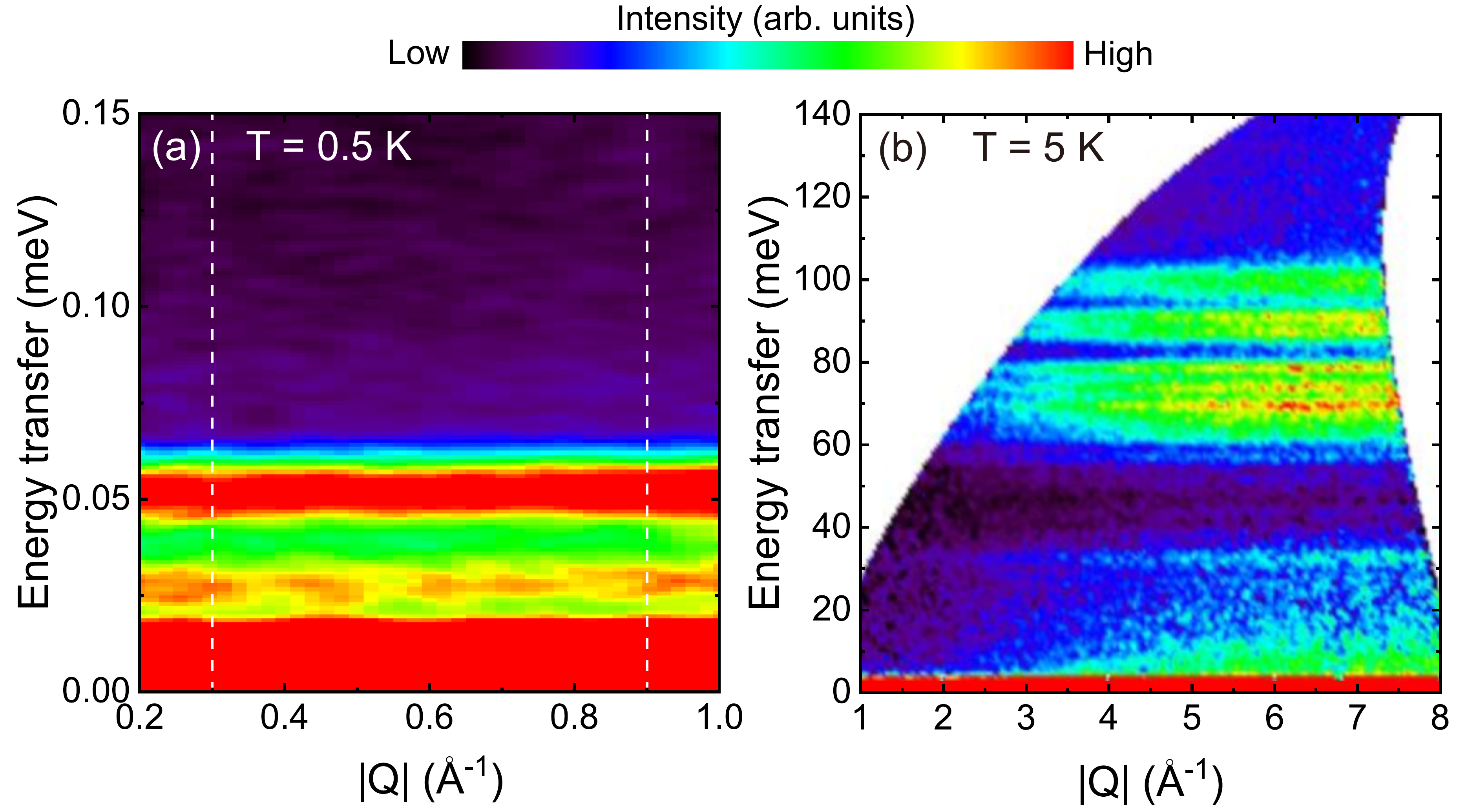}}
  \caption{~INS spectra of stepanovite~II measured at (a) CNCS and (b) SEQUOIA spectrometer. In panel (a), the region between the vertical dashed lines shows the integrated $|Q|$ range of the energy cuts, which are shown in Figs.~\ref{ins-zerofield}(c)-\ref{ins-zerofield}(f).
  }
  \label{ins-sequoia}
\end{figure}

\section{Results and discussion}

\subsection{Inelastic Neutron Scattering}\label{Sec:INS}

\begin{figure}[tb]
\center{\includegraphics[width=0.9\linewidth]{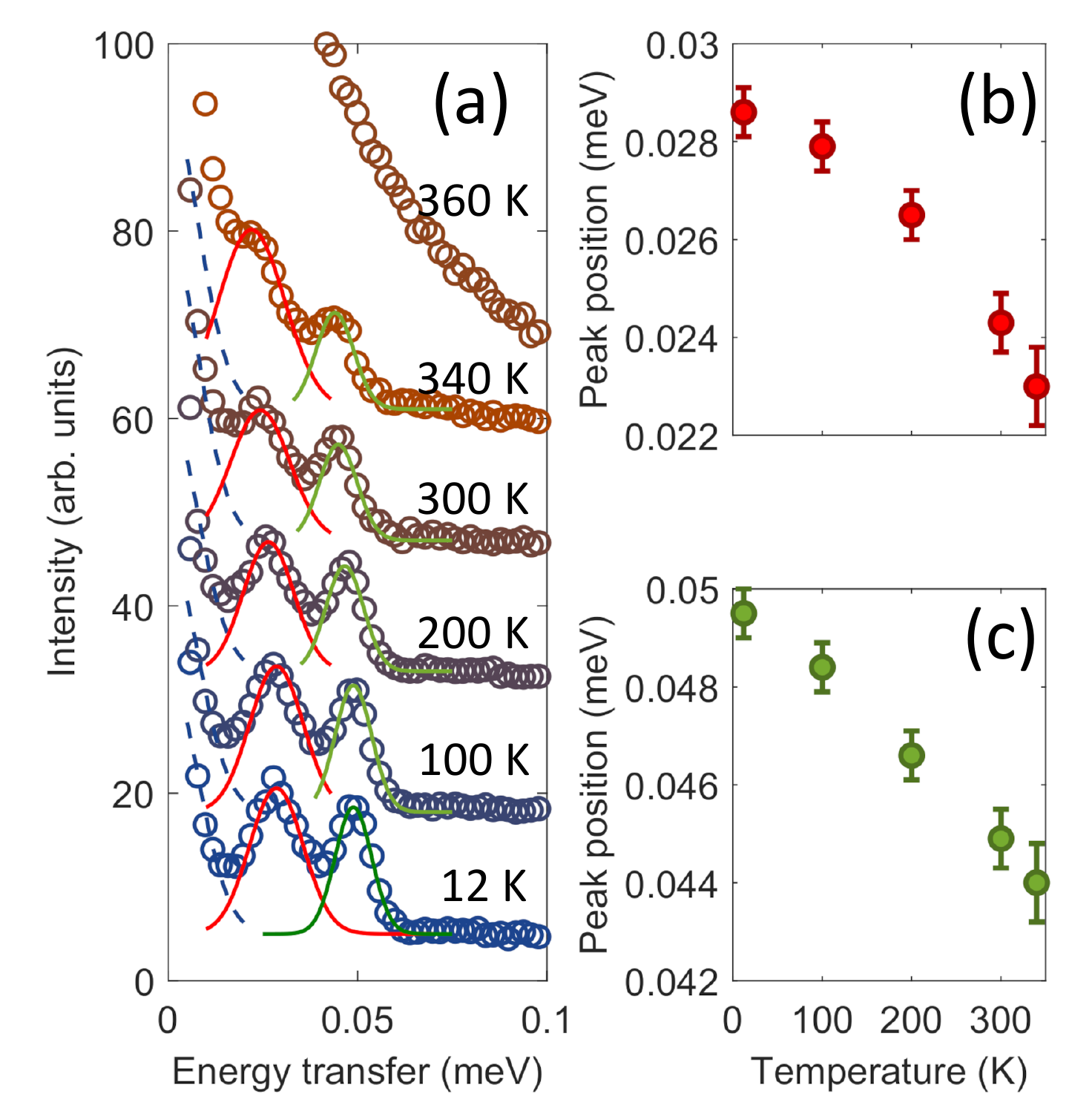}}
  \caption{~(a) INS spectra of stepanovite II measured at different temperatures at BASIS spectrometer. The integrated $|$$Q$$|$ range of these curves is $[0.2-2.0]$~\AA$^{-1}$. Red and green solid lines show deconvolution of the observed signal with two Gaussian functions and the blue dotted lines indicate background contribution due to incoherent scattering. Temperature dependence of the peak positions are summarized in panels (b) and (c). The spectra in panel (a) have been shifted vertically for clarity.
  }
  \label{ins_basis}
\end{figure}

\begin{figure*}[tb]
\center{\includegraphics[width=0.95\linewidth]{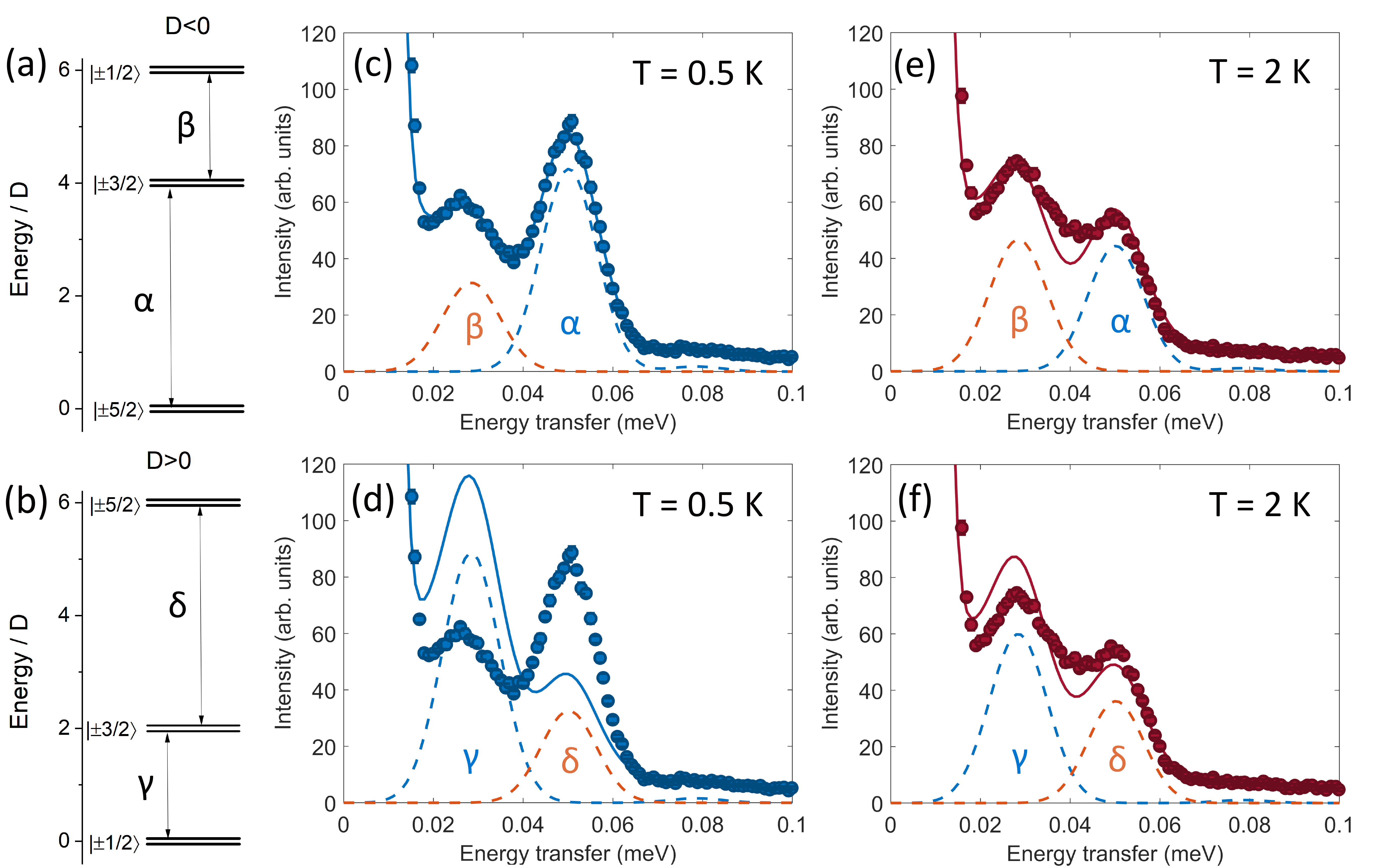}}
  \caption{~Schematic zero-field splitting and possible INS transitions of the $S$~=~5/2 for (a) negative and (b) positive $D$ parameter, and INS spectra taken at CNCS at temperatures (c,d) 0.5~K and (e,f) 2.0~K. The solid line is a fitting on top of instrumental resolution. Individual excitations are represented by a Gaussian line shape (broken lines). The integrated $|Q|$ range of the data points is $[0.3-0.9]$~\AA$^{-1}$, which is the region between the vertical broken lines in Fig.~\ref{ins-sequoia}(a).
  }
  \label{ins-zerofield}
\end{figure*}

INS experiments on a sample of polycrystalline stepanovite II were performed using three time-of-flight (TOF) spectrometers at the Spallation Neutron Source (SNS) at the Oak Ridge National Laboratory, the fine-resolution Fermi chopper spectrometer (SEQUOIA)~\cite{SEQ}, the Backscattering Spectrometer (BASIS)~\cite{BASIS}, and the Cold Neutron Chopper Spectrometer (CNCS)~\cite{CNCS1,CNCS2} (see Appendix~\ref{method} for the details of INS experiments).
These instruments cover a wide range of reciprocal and energy space, allowing the study of excitations with a broad energy range from sub-millielectron volt (meV) to electron volt (eV)~\cite{suite}.
In this case, data were obtained on both the energy- and temperature-dependence of the dynamical excitations in stepanovite II.
After careful analysis of all the data obtained from three spectrometers, we conclude that only two dispersionless excitations at very low energy transfer, $\hbar \omega$~=~0.028(1) and 0.050(1)~meV [see Figs.~\ref{ins-sequoia}(a), \ref{ins_basis}, and \ref{ins-zerofield}] have a magnetic origin.

Figure~\ref{ins-sequoia}(a) shows a typical INS spectrum of the low-energy excitations obtained from the CNCS at 0.5~K.
Two $Q$-independent excitation modes can be seen clearly.
The stronger is around 0.05 meV and the weaker is around 0.03 meV.
For the INS spectrum collected at 5~K on SEQUOIA [Fig.~\ref{ins-sequoia}(b)], we observed several peaks with intensities that increase with increasing neutron momentum transfer (Q).
This is a characteristic of vibrational modes.
Although the high-energy INS spectrum contains a great deal of information on translational and librational vibrations of structural  water molecules, this work aims to clarify the magnetic properties of stepanovite II and, therefore, we will focus on the low energy part of the INS spectra in this paper.

Figure~\ref{ins_basis}(a) shows the low-energy INS spectra of stepanovite~II obtained from the BASIS spectrometer at several temperatures.
Two inelastic peaks can be observed around $\hbar \omega$~=~0.028(1) and 0.050(1)~meV at 12 K.
The intensities of both are almost temperature independent in the range of $12$~K$<T<340$~K.
This is because the thermal population of the ground and excited states does not change significantly with temperature when $k_{\rm B}T >> \Delta$, where $\Delta$ is the energy gap between the ground and excited levels (see below).
The INS peaks observed on BASIS only disappear at temperatures above the melting point $T_m \sim 350$~K of \ste.
The results show a slight broadening and shift towards low energies of the peaks with increasing temperature.
We fit the INS spectra with two Gaussian functions and extracted the evolution of the peak positions with increasing temperature [see Figs.~\ref{ins_basis}(b) and \ref{ins_basis}(c)].
The peak shifts turn out to be about 20\% and 12\% for the $\hbar \omega$~=~0.028 and 0.050~meV excitations, respectively.
We conclude that the magnetic anisotropy is decreased with increasing temperature.
The line broadening is most probably caused by local structural disorder, which increases with increasing temperature.
To clarify this question, further structural studies are necessary.

In stepanovite~II, the  Fe$^{3+}$  ions  with  a  3$d^5$  electronic  configuration are in octahedral environment,  which generates a singlet $^6$S$_{5/2}$ Hund’s  ground  state  term.
The ground state has no spin-orbit splitting, since the orbital moment is zero.
Therefore, the Fe$^{3+}$ ion has a spin-only ground state $S$~=~5/2.
The effective spin Hamiltonian can be described as
\begin{eqnarray}
 \hat{\mathcal{H}} =&& D(\hat{S}^2_z - \frac{1}{3}S(S+1)) + E(\hat{S}^2_x - \hat{S}^2_y)
 \nonumber\\
&&+ \mu_{\mathrm{B}} \bm{g} \bm{B} \bm{\hat{S}},
 \label{Ham}
\end{eqnarray}
where $D$ and $E$ are the axial and rhombic ZFS parameters, $\bm{\hat{S}} = (\hat{S}_x,\hat{S}_y,\hat{S}_z)$ is the $S$~=~5/2 spin operator, $\mu_{\mathrm{B}}$ is Bohr magneton, $\bm{g}$ is the Land$\acute{e}$ $g$-factor and $\bm{B}$ is the applied magnetic field~\cite{Suzuki,Mossin}.
A negative $D$ corresponds to an easy-axis anisotropy (out-of-plane anisotropy), while a positive $D$ relates to an easy-plane anisotropy~\cite{Novitchi}.
In the absence of magnetic field, the $S$~=~5/2 sextet is split into three Kramers doublets, $|\pm 1/2 \rangle$, $|\pm 3/2 \rangle$ and $|\pm 5/2 \rangle$.
In the case of zero rhombic distortion, the energy splittings between the three doublets are 2$D$, 4$D$ and 6$D$ [see Figs.~\ref{ins-zerofield}(a) and \ref{ins-zerofield}(b)]~\cite{Febbraro}.
Our experimental data shows slight deviation of the positions of the Kramers doublets from the ideal case.
Besides, we observed a weak INS intensity at $\sim 0.078$~meV, which corresponds to the transition from the ground to second exited state (Fig.~\ref{ins-zerofield}).
Thus, the rhombic ZFS parameter has nonzero value.

In order to determine the values of the $D$ and $E$ parameters in the Hamiltonian (\ref{Ham}), and especially the sign of $D$, we analyzed the results obtained at $T=0.5$~K, where thermal populations of the ground state and excited levels differ considerably, and compared the result with that at $T=2$~K (see Fig.~\ref{ins-zerofield}).
Since the intensity of the transition from the ground state increases with the decreasing temperature, comparison of the relative peak intensities of the INS spectra at $T=0.5$~K and $T=2$~K will unambiguously show which Kramers doublet, $|\pm 1/2 \rangle$ or $|\pm 5/2 \rangle$, is the lowest energy one.
To do so, the INS spectra were simulated by calculating the energies $E_n$ and corresponding wave functions $|n \rangle$ via exact diagonalization of the Hamiltonian (\ref{Ham}).
For experiments on polycrystalline materials, the INS intensity of the transition from level $i$ to level $f$ is proportional to the magnetic scattering function as~\cite{Basler}
\begin{eqnarray}
\mathcal{S}(\hbar \omega) &\propto& N p_i \sum_{\alpha}| \langle f | \hat{S}_{\alpha} | i \rangle |^2
\nonumber \\
&&\times P(\hbar \omega - \Delta_{i\rightarrow{}f}, \mathrm{FWHM}_{i\rightarrow{}f}),
\label{S}
\end{eqnarray}
where $N$ is the total number of spins, $\alpha = x,~y,~z$, $\hbar \omega$ is the neutron energy transfer, $| i \rangle$, and $| f \rangle$ are spin eigenstates with energy $E_{i}$, $E_{f}$, and $p_i$ is the Boltzmann population factor.
$P(\hbar \omega - \Delta_{i\rightarrow{}f}, \mathrm{FWHM}_{i\rightarrow{}f})$ is the line shape of the peak centered at energy transfer $\Delta_{i\rightarrow{}f}$, and FWHM is the full width at half maximum of each peak.

In Figs.~\ref{ins-zerofield}(c)-\ref{ins-zerofield}(f), we present the experimental INS spectra, together with the calculated spectra for both the $D$ $\textless$ 0 [(c) and (e)] and $D$ $\textgreater$ 0 [(d) and (f)] cases.
Transitions $\alpha$ and $\gamma$ correspond to excitations from the ground state to the first excited state, while transitions $\beta$ and $\delta$ are excitations between the thermally populated first excited state and the higher energy second excited state.
The  excitations are fitted by two Gaussian functions with corresponding energy dependent resolution~\cite{CNCS1}.
One can see that the agreement between calculation and experiment is much better for the negative $D$ parameter, meaning that the Fe$^{3+}$ ion has an easy-axis anisotropy with $|\pm 5/2 \rangle$ as the ground state.
The best fitting for both temperatures, 0.5 and 2~K, was found using the values of $D$~$=-0.0128(5)$~meV and $E$~$=0.0014(5)$~meV.

\subsection{Heat capacity}

\begin{figure}[t!]
\center{\includegraphics[width=1\linewidth]{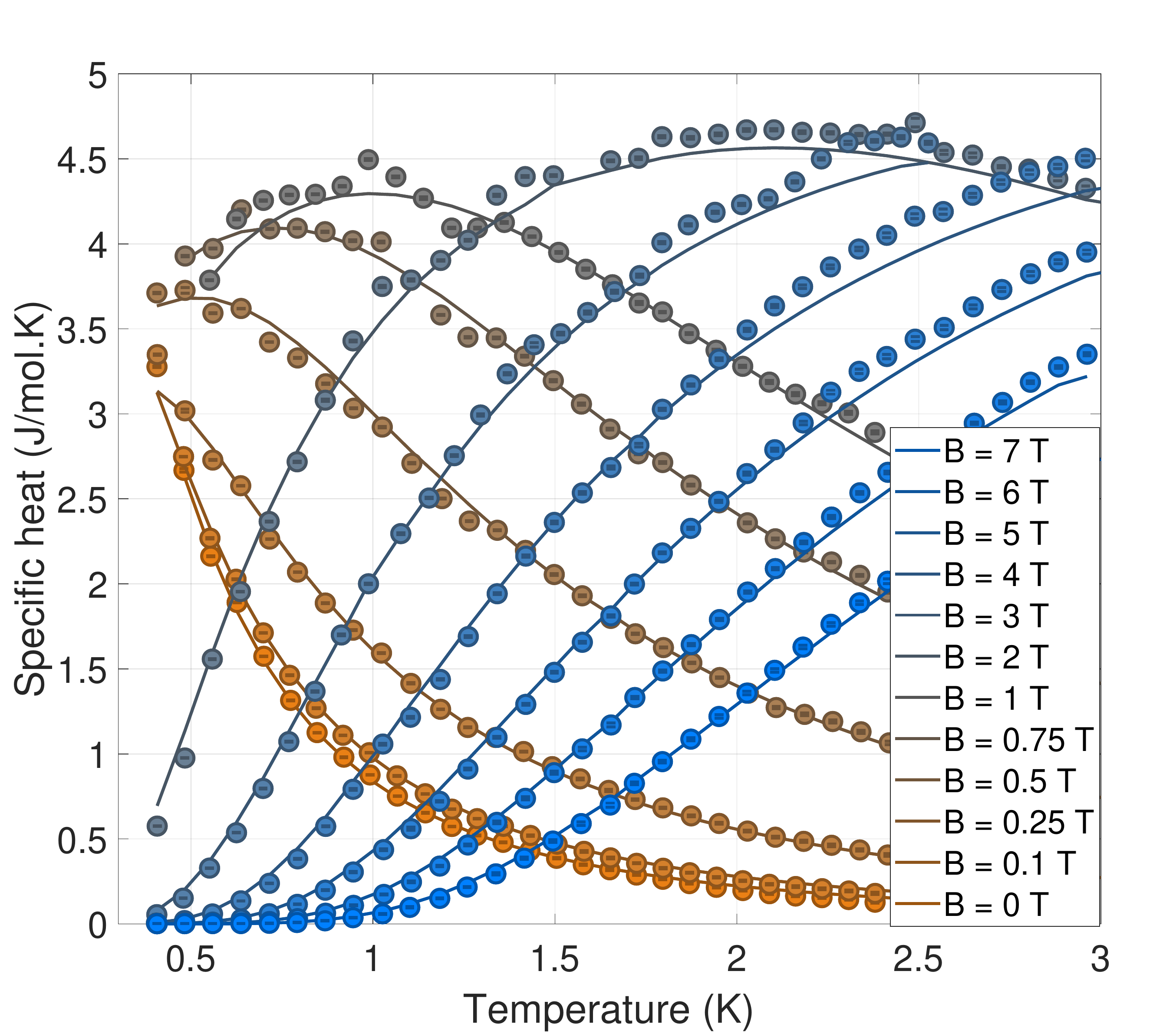}}
  \caption{~Specific heat as a function of temperature measured at different magnetic fields. Dots show experimental data, lines represent the calculated curves as described in the main text. Note that a small phonon contribution ($\sim0.4$~J/mol.K at 3~K) was fitted with $C \propto \alpha\cdot T^3$ and subtracted from the data to isolate the magnetic signal.
  }
  \label{Fig:Specific_heat}
\end{figure}

The specific heat is a useful thermodynamic probe, because it is sensitive to  low-energy degrees of freedom.
For instance, the specific heat of a two-level system exhibits a Schottky anomaly due to thermal population of the excited state, which can be used to precisely determine the splitting energy.
For a $n$-level system the expression for the Schottky anomaly can be rewritten as:
\begin{eqnarray}
C =  R \frac{d}{dT} \Big( \frac{1}{\mathcal{Z}} \sum_{i = 1}^n \ E_ie^{-\frac{E_i}{k_{\mathrm{B}} T} }\Big),
 \label{MultiSchottky}
\end{eqnarray}
where $R$ is the universal gas constant, $\{E_i\}$ is a set of energy levels ($n = 2S + 1 = 6$), and $\mathcal{Z}$ is the partition function.

In order to characterize low-energy excitations of stepanovite~II, we performed a single-crystal specific heat measurements at different magnetic fields.
The results are summarized in Fig.~\ref{Fig:Specific_heat}~\footnote{Note that the weak phonon contribution was modeled with $C \propto \alpha\cdot{}T^3$ and subtracted from the data.}.
No sign of any phase transitions can be observed down to 0.5~K with and without a field present.
At zero field, the specific heat shows a strong increase with decreasing temperature.
Application of magnetic field induces a broad, Schottky-like peak, which continuously evolves with the increasing field strength.

\begin{figure*}[t!]
\center{\includegraphics[width=0.85\linewidth]{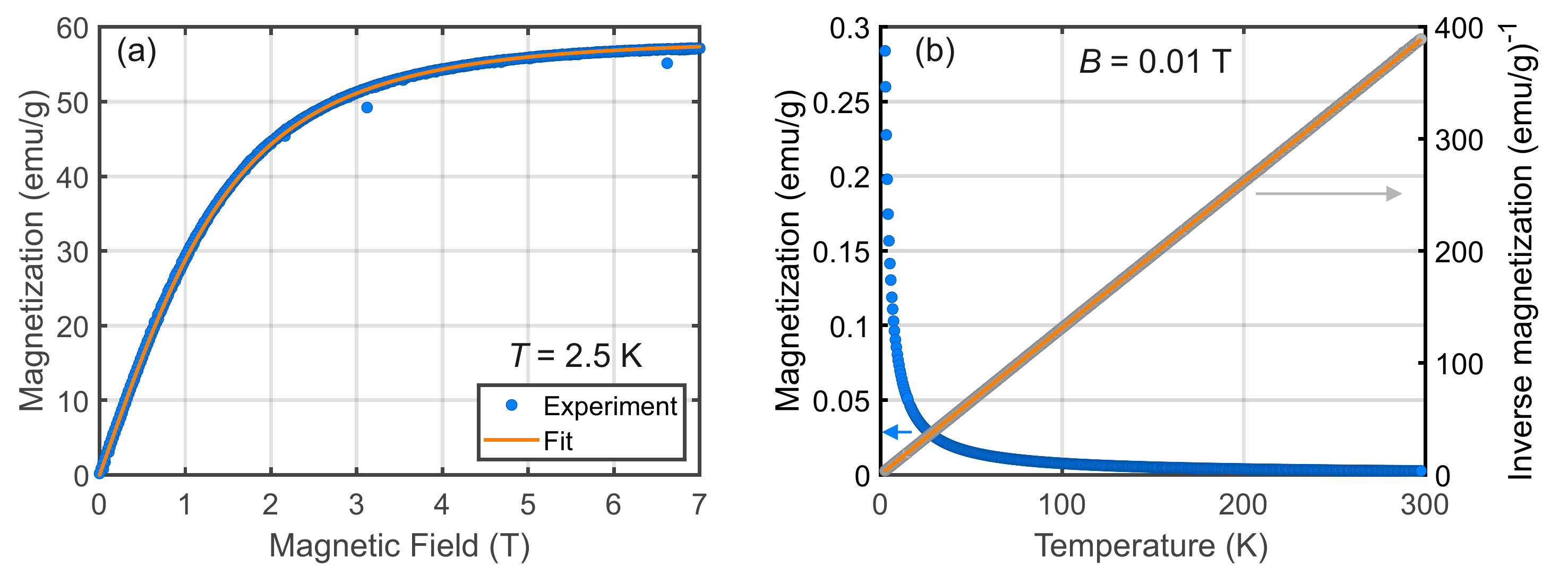}}
  \caption{~(a) Magnetic field dependence of magnetization measured at $T = 2.5$~K with $B \perp c$ and its fit with Eq.~\eqref{Eq:magnetization}. (b)~Temperature dependencies of magnetization and inverse magnetization obtained at $B = 0.01$~T applied along the same direction as in (a).
  }
  \label{Fig:Magnetization}
\end{figure*}

To describe the observed curves, we diagonalized the Hamiltonian~\eqref{Ham}.
The resultant eigenenergies were substituted into Eq.~\eqref{MultiSchottky} (we assumed $g = 2$ and $B \parallel  y$).
We then performed a global fitting of E and D parameters using all the data obtained.
The resulting CEF parameters were found to be $D$ = $-0.0136(4)$~meV, $E$ = 0.0013(5)~meV.
These values are almost identical to those deduced from the INS measurements in Sec.~\ref{Sec:INS}.
The calculated $C(T)$ curves are shown as solid lines in Fig.~\ref{Fig:Specific_heat}.
A near perfect agreement was obtained between the calculated and experimental curves, although we had to multiply the calculated curves by a constant scaling prefactor (0.713).
This maybe related to the presence of impurities in the sample.
This further confirms our previous conclusions concerning the magnetic nature of the excitations around 0.03 meV and 0.05 meV in INS spectra.

\subsection{Magnetization}

To further characterize the magnetic properties of stepanovite~II, we performed measurements of static magnetization on single crystal samples.
Figure~\ref{Fig:Magnetization}(a) shows the field dependence of the magnetization measured at $T = 2.5$~K.
The curve exhibits a Brillouin-like behavior.
To quantify this result we diagonalized the Hamiltonian~\eqref{Ham} at each field $H_y$ and calculated the magnetization as:
\begin{align}
    M_{y} = g\mu_{\rm B} \frac{\sum_{i = 1}^n \langle i | \hat{S}_{y} | i\rangle e^{-E_i/k_{\mathrm{B}}T}}{\sum_{i = 1}^n e^{-E_i/k_{\mathrm{B}}T}},
    \label{Eq:magnetization}
\end{align}
where $|i\rangle$ and $E_{i}$ are the eigenvector and eigenvalue of the Hamiltonian, respectively~\footnote{We multiplied the calculated curve by constant prefactor to fit the experimental magnetization scale.}.
The calculated curve is shown in Fig.~\ref{Fig:Magnetization}(a) as a solid line. This agrees well with the experimental data.
However, the experimental temperature $k_{\rm B}T = 0.215$~meV $\gg |D|=0.0128$~meV.
Therefore, the influence of the CEF term is rather weak and the simple Brillouin function for $S = 5/2$ with $D$~=~0 also provides a reasonable description of the measured magnetization.

The temperature dependence of the magnetization and inverse magnetization are presented in Fig.~\ref{Fig:Magnetization}(b).
The magnetization data display perfect Curie-Weiss behavior.
The inverse magnetization is linear over the whole temperature range down to 2~K. Its fitting yields a very small positive Curie-Weiss temperature $\Theta_{\rm CW}= 0.2$~K, which indicates a very weak dominant ferromagnetic coupling between Fe$^{3+}$ ions persistent down to low temperatures.
The static magnetization, therefore, shows that stepanovite II is paramagnetic.

\section{Conclusion}

In conclusion, we have presented INS, magnetization and heat capacity studies of Fe$^{3+}$ ions in the oxalate mineral \ste.
The INS spectra show two excitation modes from magnetic transitions between ZFS of $S$~$=5/2$ ground state multiplets of Fe$^{3+}$ ions.
The results reveal an easy-axis anisotropy of Fe$^{3+}$ in the stepanovite II, with zero-field parameters $D$~$=-0.0128(5)$~meV and $E$~$=0.0014(5)$~meV, which have been further confirmed by the analysis of the heat capacity results.
There is no magnetic order down to 0.5~K, which makes the material a pure paramagnet.
Note that MOF materials are soft porous crystals displaying large-amplitude structural deformations under external physical constraints such as pressure.
Therefore, magnetically ordered ground states due to increased magnetic exchange coupling can occur under appropriate pressures.
While no high-pressure studies of \ste\ have been reported, we hope that our work will inspire further studies of this quasi-2D MOF material.

\section*{Acknowledgments}
Work at Oak Ridge National Laboratory was supported by the U.S. Department of Energy (DOE), Office of Science, Basic Energy Sciences, Materials Science and Engineering Division.
Work by LMA was supported by the U.S. Department of Energy, Office of Science, Office of Basic Energy Sciences, Chemical Sciences, Geosciences, and Biosciences Division.
We would like to thank D.M.~Pajerowski and I.A.~Zaliznyak for stimulating discussions and J. Yan for the help with magnetic measurements.
This research used resources at the Spallation Neutron Source, a DOE Office of Science User Facility operated by the Oak Ridge National Laboratory.
X-ray Laue measurement was conducted at the Center for Nanophase Materials Sciences (CNMS) (CNMS2019-R18) at the Oak Ridge National Laboratory (ORNL), which is a DOE Office of Science User Facility.

This paper has been authored by UT-Battelle, LLC under Contract No. DE-AC05-00OR22725 with the U.S. Department of Energy. The United States Government retains and the publisher, by accepting the article for publication, acknowledges that the United States Government retains a non-exclusive, paid-up, irrevocable, world-wide license to publish or reproduce the published form of this manuscript, or allow others to do so, for United States Government purposes. The Department of Energy will provide public access to these results of federally sponsored research in accordance with the DOE Public Access Plan (http://energy.gov/downloads/doe-public-access-plan).
\appendix
\section{Technical Details}
\label{method}

Synthetic stepanovite II \ste\ was obtained by reacting Fe$_2$O$_3$ and MgO
with aqueous NaOH and oxalic acid \cite{Huskic2016}.
After 2 days, the solution yielded green crystals with a trigonal habit about a millimeter across.

The INS data were collected from a polycrystalline sample with a mass of around 3.0~g using the BASIS, SEQUOIA and CNCS TOF spectrometers \cite{suite}.
The incident neutron energy for every measurement was chosen to cover the anticipated region of interest in both energy transfer $\hbar \omega$ and scattering-vector $Q$ spaces.
A low incident energy $E_i$ is especially important to observe excitations close to the elastic peak as the energy resolution is typically $1.5 - 2$\% of the incident energy.
For the SEQUOIA and BASIS measurements the sample was mounted in a standard closed cycle refrigerator (CCR) with a
base temperature of $T \sim 5$~K.
A standard Orange cryostat with $^3$He-refrigerator insert was used at the CNCS in order to expand the low temperature range to $\sim 0.5$~K.
The $E_i$ was fixed at 800, 600, 250, 160 and 55~meV (SEQUOIA), and $E_i$~=~1.55~meV (CNCS).
The $|Q|$ range was 0.2 $-$ 2.0~\AA $^{-1}$ for BASIS and 0.1 $-$ 1.1~\AA $^{-1}$ for CNCS.
Vanadium was used as a standard for the detector efficiency correction.

The thermodynamic measurements were performed using single-crystalline samples with masses of $\sim$2 $-$ 4~mg and the applied magnetic field was perpendicular to the $c$-axis.
Magnetization was measured using a vibrating sample magnetometer MPMS SQUID VSM, Quantum Design in the temperature range 2 $-$ 300~K.
Specific-heat measurements were carried out using a commercial PPMS-6000 instrument  from Quantum Design equipped with $^3$He insert to reach temperatures as low as 0.4~K

The software packages \textsc{Dave}~\cite{Dave}, and \textsc{MantidPlot}~\cite{Mantid} were used for data reduction and analysis.

\bibliography{stepanovite}

\end{document}